\documentclass[apjl]{emulateapj}
\usepackage{hyperref}
\usepackage{color}
\hypersetup{colorlinks,
 citecolor=blue,
 linkcolor=blue}
\usepackage{soul}
\usepackage{natbib}

\usepackage{hyperref}
\hypersetup{colorlinks,%
 citecolor=blue,%
 linkcolor=blue}
\definecolor{lightblue}{rgb}{.70,.95,1} 
\def\aj{AJ}
\def\apj{ApJ}
\def\apjl{ApJ}
\def\apjs{ApJS}
\def\aap{A\&A}
\def\mnras{MNRAS}
\def\araa{ARA\&A}

\slugcomment{Draft Version \today}
\shorttitle{UGC 4703 pair: an LMC-SMC analog.}
\shortauthors{Paudel et al.}
\begin{document}
\title{UGC 4703 interacting pair near to an isolated spiral galaxy NGC 2718: A Milky Way Magellanic Cloud analogue} 

\author{Sanjaya Paudel\altaffilmark{1}, C. Sengupta \altaffilmark{2}
  }
\affil{
$^1$ Department of Astronomy \& Center for Galaxy Evolution Research, Yonsei University, Seoul 03722, Korea\\
$^2$ Department of Astronomy, Yonsei University, 50 Yonsei-ro, Seodaemun-gu, Seoul, Republic of Korea\\
}

\altaffiltext{2}{Email: sanjpaudel@gmail.com}

\begin{abstract}
We present an analysis of physical and morphological properties of an interacting pair of dwarf galaxies UGC 4703 located at the vicinity of an isolated Milky-Way (MW) type spiral galaxy NGC 2718. Based on the comparison of physical and morphological properties with that of the Large and Small Magellanic Clouds (LMC and SMC), we report that the UGC 4703 pair-NGC 2718 system is probably an LMC-SMC-MW analogue. Located at a sky projected distance of 81 kpc from NGC 2718, we find that UGC 4703 is clearly interacting with its nearby smaller mass companion UGC 4703B forming a bridge of stellar stream between them. Total B-band luminosity of UGC 4703 and its companion is -17.75 and -16.25 mag, respectively.   We obtained HI 21-cm  line data of UGC 4703 using the GMRT  to get a more detailed view of neutral hydrogen (HI) emission. The HI image revealed evidence of interaction between the dwarf galaxy pair but no extended emission, like Magellanic Stream (MS),  was detected between NGC 2781 and the UGC 4703 dwarf pair. We also detected star-forming regions along the UGC 4703/ 4703B bridge with stellar mass exceeding 10$^{7}$ M$_{\sun}$. While comparing the optical and HI morphology of the interacting dwarf pairs (UGC 4703-4703B and LMC-SMC) we discuss  possible difference in interaction histories of these systems.

\end{abstract}


\section{Introduction}
Unusual morphology of the Large and Small Magellanic clouds (LMC and SMC) in optical and radio observations has been used to interpret their close encounter \citep{Gardiner96},  and the resultant triggered star formation \citep{Harris09,Glatt10}. \cite{Besla07} suggested that both dwarf galaxies may be entering the  Milky-Way (MW)  system as a binary pair for the first time. Magellanic clouds are the only bright and star-forming satellites of MW. Number of studies have shown that a satellite pair of  LMC-SMC mass around MW mass host is  neither common in observation, nor in numerical simulation with hierarchical accretion \citep{Boylan-Kolchin10,Robotham12,Tollerud11}. Analyzing a result of the Millennium-II Simulation, \cite{Boylan-Kolchin10} predicted that there is a less than 10 per cent chance that MW mass halo hosts two subhalos of mass of Magellanic clouds. Similarly, analyzing the Galaxy And Mass Assembly (GAMA) catalog of galaxies, \cite{Robotham12} calculated the probability of such system is less than 5\%.  Following this statistics, the merger probability of LMC-SMC morphology dwarf galaxy satellites around the MW mass host maybe even smaller.
 
It is commonly believed that, by having a shallow potential well, low-mass dwarf satellites should be more influenced by tidal potential of parent halos, and less so by the merging events \citep{Paudel13,Paudel14b}. Nevertheless, mounting evidence suggests that merger of dwarf galaxies might not be as rare as it was previously thought. Apart from the classic LMC-SMC interaction, there are several cases of merging dwarf galaxy candidates reported in recent studies \citep{Delgado12,Johnson13,Crnojevic14,Paudel15,Stierwalt15}. In our previous work, we studied merging of gas-rich dwarf galaxies, UGC 6741 pair, which is located in a low density environment, i.e outskirt of the NGC 3853 group  \citep{Paudel15}.  We found UGC 6741 has a similar visual morphology of ARP 104.

\begin{figure*}

\includegraphics[width=18cm]{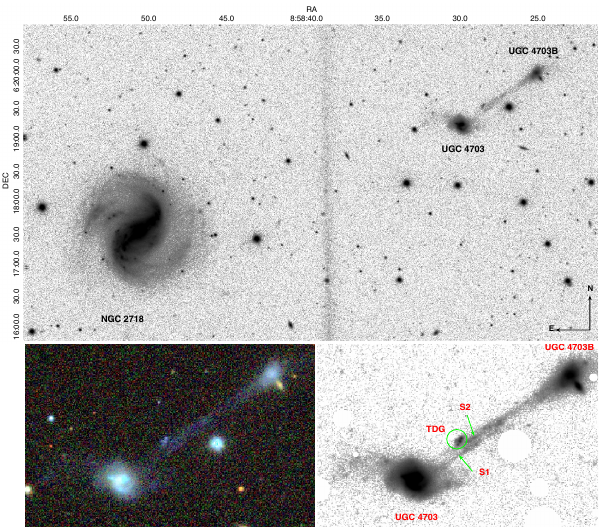}
\caption{We show the SDSS $r-$band image of an area around NGC 2718 and its satellites with a field of view ofRain 9$\arcmin$$\times$5$\arcmin$. In the bottom panel, we show a zoom in view of interacting dwarf galaxies. The left is a color image cutout directly obtained from the SDSS sky-server.  Right, we show co-added $g$, $r$ and $i$-band images with an arcsinh scaling.  Some interesting features, like S1 and S2 and TDG, are highlighted.}
\label{mainfigure}
\end{figure*}

Here we report one  more system of merging dwarf galaxies near to the MW type isolated host galaxy, NGC 2718. This resembles the LMC-SMC-MW system in various ways. This paper is organized as follows. In Section 2, we introduce the system and the environment. Section 3 describes the photometric characterization and morphology, while Section 4 is dedicated to discussing its importance.

\section{Identification}
As our primary interest is to perform a detailed study of merging system of dwarf galaxies in various environments, we carried out a systematic search of such objects visually inspecting the SDSS color image in the local volume (z $<$ 0.02). In this work, we present UGC 4703,  an interacting pair of dwarf galaxies. It is  similar to our previously studied galaxy UGC 6741, but located near MW mass spiral galaxy NGC 2718. The main aim of this work is to report the similarity between the UGC 4703 interacting pair around NGC 2718 and the LMC-SMC interaction around MW.

At the position RA=08:58:29.75, DEC=+06:19:16.8, and a redshift of z = 0.01199, we found a rare pair of star-forming dwarf galaxies with a connecting bridge of stellar stream. The brighter galaxy in the pair,  UGC 4703, is slightly fainter than LMC with an $r-$band absolute magnitude M$_{r}$ = -18.25 mag and the fainter companion, here after UGC 4703B, has an $r-$band absolute magnitude M$_{r}$ = -16.76 mag. The pair is located North-West of  NGC 2718, at an angular distance of $\approx$5.2$\arcmin$. Assuming the distance to the NGC 2718 group to be 54.5 Mpc\footnote{Based on Hubble flow with a redshift z= 0.012}, the physical projected separation between  NGC 2718 and UGC 4703 is 81 kpc. A relative line of sight redial velocity between the two is 263 km/s. According to available notes in the NED\footnote{http://ned.ipac.caltech.edu}, UGC 4703 is already identified as a star-forming pair of galaxies interconnected by a thin straight bridge.

\section{Data analysis}\label{data}
This work benefits from a substantial multi-wavelength data available in public archives. This allowed us to perform the required detailed analysis of morphology and stellar population properties of our sources. We report a multi-wavelength study of the system based on the archival images from the Sloan Digital Survey (SDSS), Galaxy Evolutionary Explorer (GALEX) and the Spitzer space telescope which covers a wavelength range from Far Ultraviolet to Infra Red. Additionally, we observed the system in HI 21-cm line using the Giant Metrewave Radio Telescope (GMRT). 

\begin{table*}
\label{aphot}
\caption{Results of aperture photometry}
\begin{tabular}{cccccccccccccc}
\hline
 Galaxy & Ra & Dec & FUV & NUV & u & g & r & i & z & [3.6] & [4.5]  & log(SFR)& log(M$_{*}$)\\
 & h:m:s & d:m:s & mag &  mag &  mag &  mag &  mag &  mag &  mag & mJy & mJy  & M$_{\sun}$/yr & M$_{\sun}$\\
\hline 
 UGC 4703 & 08:58:29.869 & 06:19:17.07  & 17.1  & 16.8  & 16.31   & 15.60   & 15.26   & 15.19   & 15.08 & 2.060 & 1.540 & -0.64 & 9.27\\
 UGC 4703B& 08:58:25.024 & 06:20:06.44  & 19.6  & 18.9  & 18.25   & 17.10   & 16.76   & 16.50   & 16.38 & 0.390 & 0.230 & -1.49 & 8.74\\
 TDG      & 08:58:28.424 & 06:19:34.76  & 20.4  & 20.3  & 20.05   & 19.25   & 19.32   & 19.20   & 19.30 & 0.010 & 0.005 & -2.05 & 7.28\\
\hline
\end{tabular}
\\
The UV magnitudes are from the GALEX images and the optical are from the SDSS. Spitzer space telescope images are used to derive IR flux. UV and Optical magnitudes are corrected for galactic extinction. We used \cite{Schlafly11} extinction map and we calculated the extinction in UV using \cite{Seibert05}.The stellar mass are derived from Spitzer [3.6] and [4.5] channel flux and star-formation rates are derived from the GALEX FUV flux.
\end{table*}

\subsection{Analysis of archival data}\label{analysis}

In Figure \ref{mainfigure}, we show the SDSS $r-$band image of an area around NGC 2718 and its satellites with a field of view of 9$\arcmin$$\times$5$\arcmin$. NGC 2718  is a well studied isolated galaxy \citep{Hernandez10,Karachentseva10}. We   find no bright galaxy (M$_{r}$ $<$ -19 mag) around it within one Mpc sky-projected radius and a radial velocity $\pm$500 km/s. Our NED query found no other dwarf  (M$_{r}$ $>$ -19 mag) companion other than the two, UGC 4703  and UGC 4703B, around NGC 2718. NGC 2718 is a nuclear star-burst galaxy and in RC3 catalog, it is classified as a face-on SAB type. It has a total stellar mass of Log(M$_{*}$/M$_{\sun}$) = 10.7 and total HI mass of Log(M$_{HI}$/M$_{\sun}$) = 10.05 \citep{Chang15,Courtois15}. Only, NGC 2718 and UGC 4703 were targeted by the SDSS spectroscopy survey, and measured redshift for them are 0.0127 and 0.0119, respectively. Unfortunately, the SDSS target selection picks up a background red galaxy located on the edge of UGC 4703B, however the NED cataloged redshift of UGC 4703B is 0.0118.

A stellar bridge connecting UGC 4703 and its companion UGC 4703B can be clearly seen in the SDSS image. A sky-projected separation between the two is 20 kpc. In Figure \ref{mainfigure} lower panel, we show a more detailed view of the interacting pair. In the left, we show a color image cutout directly obtained from the SDSS sky-server\footnote{http://skyserver.sdss.org}.  In the $g-r-i$ combined color image, we can see a few extremely blue knots in the central region of UGC 4703 which indicate ongoing burst of star-formation. The H$_{\alpha}$ emission line equivalent width measured from the SDSS optical spectrum of UGC 4703 is 181 \AA{}. This gives a star-formation age of order of a few Myr \citep{Leitherer99}. The optical morphology of UGC 4703B however  is much smooth where no distinct blue star-forming knots is visible. From a detailed inspection of UGC 4703 morphology in the Figure \ref{mainfigure}, bottom right panel, it seems like a spiral galaxy and now one of it's spiral arms is outflung in the direction of NGC 2718.

The stellar bridge connecting the interacting pair is not homogeneous in surface brightness and color, and seems bifurcated (see Figure \ref{mainfigure} lower right panel for more detail). Interestingly, each one of these streams (S1 and S2) seems to originate from different galaxies. It is impossible to reckon the difference between these two in terms of stellar population properties from the shallow SDSS images, but the stellar bridge, overall, is bluer than the both interacting galaxies. We find a bright star-forming clump nearly at the end of the stellar stream starting from UGC 4703B, i.e S2. It may be a potential candidate of Tidal Dwarf Galaxy (TDG) in formation\footnote{Exact definition of TDG is vague. Here, a working definition is any stellar system of mass range $\gtrsim$ 10$^{7}$ M$_{*}$/M$_{\sun}$ born out of expelled debris of interacting galaxies.} In the left color image, we can see that it is much bluer than the surrounding stellar stream.

We performed the aperture photometry to derive total brightness of the objects of interest, i.e  UGC 4703, UGC 4703B and the potential TDG candidate. The sizes of apertures were selected visually where we used a wide enough aperture that secure all the flux. We used a similar approach to subtract the sky-background count as in \cite{Paudel15}. Before the doing aperture photometry, we masked all unrelated foreground and background objects manually.  The GALEX all-sky survey \citep{Martin05} archival images were used to derive the UV magnitudes.  We used the  SDSS-III \citep{Abazajian09} image to derive the optical band magnitudes. IR fluxes were measured from the Spitzer space telescope which were obtained from IRSA\footnote{http://irsa.ipac.caltech.edu} archive. We list the results of aperture photometry in various band from UV to IR in Table \ref{aphot}.

We find the both the dwarfs, UGC 4703 and UGC 4703B, have similar $g-r$ color indices of 0.34 mag. The putative TDG $g-r$ color index is -0.07 mag whereas an average $g-r$ color index of the stellar bridge is 0.1 mag. We derived the star-formation rate from the FUV flux using a calibration provided by \cite{Kennicutt98}. \cite{Eskew12} calibration was used to convert the Spitzer [3.6] and [4.5] channel flux to stellar mass. The total $r-$band luminosity  and stellar mass of UGC 4703, UGC 4703B and TDG candidate were estimated to be -18.25, -16.75 and -14.19 mag and 1.9$\times$10$^{9}$, 5.5$\times$10$^{8}$ and 1.9$\times$10$^{7}$ M$_{\sun}$, respectively.

\subsection{Radio 21-cm observation}\label{radio}

\begin{figure*}
\label{main}
\includegraphics[width=18cm]{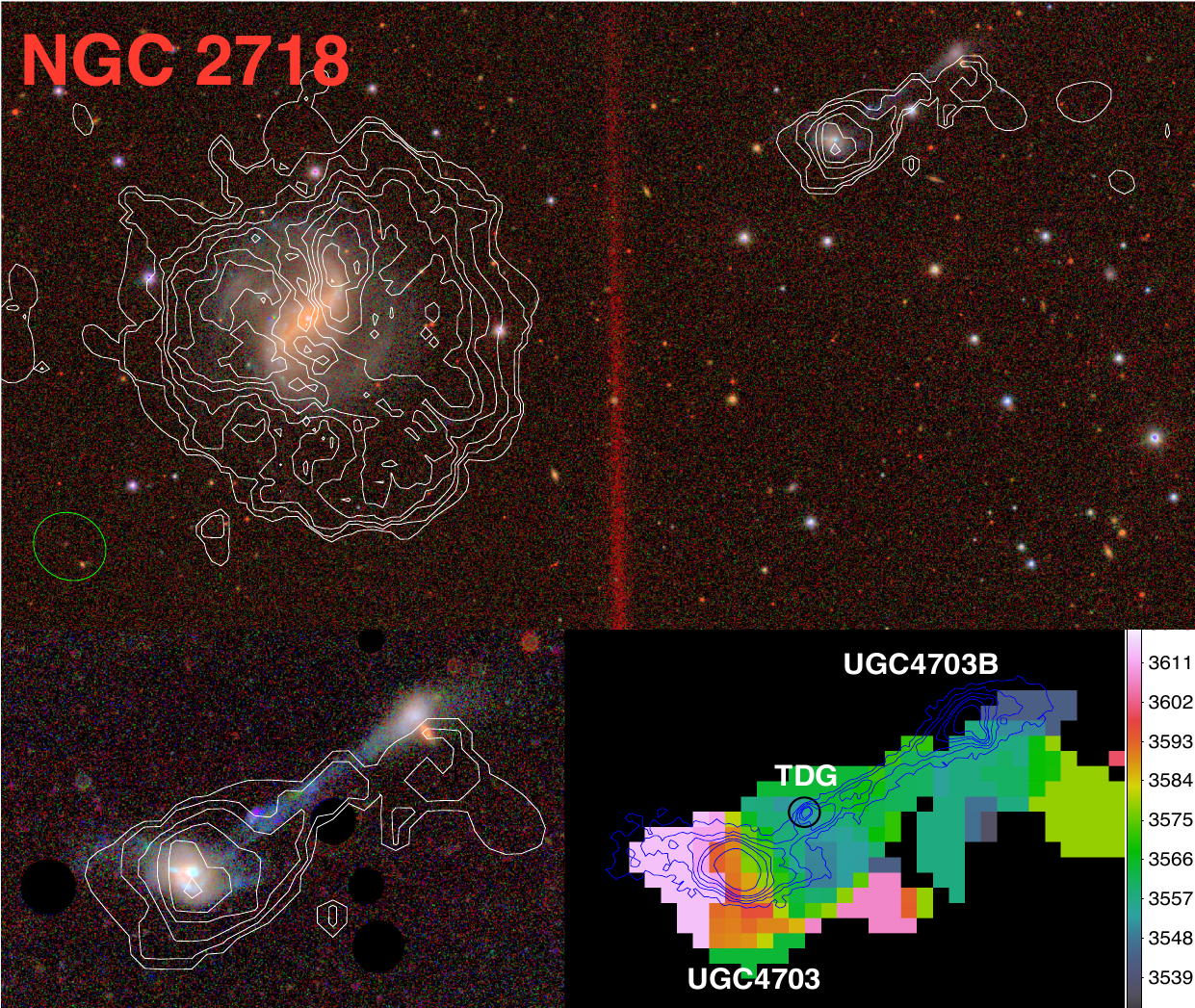}
\caption{Top: integrated HI contours from the GMRT low-resolution map overlaid on the SDSS g-r-i combined tricolor image. The H i column density levels are N(HI) = 10$^{19}$ (2.4, 4.1, 6.6, 9.9, 13.3, 16.6, 19.9, 23.3, 26.7, 30.0)   cm$^{-2}$\\
Bottom: we show a zoom-in view of area around UGC 4703 and its companion. In right, we show HI velocity field from the GMRT low-resolution cube for HI emission $>$3$\sigma$. The blue contour traces the SDSS r-band light where the outermost is of surface brightness 24.5 mag/arsec$^{2}$. The black circle represents position of TDG. }
\label{hiful}
\end{figure*}

  To get a detailed view of the HI distribution of the UGC 4703 system and its connection to NGC 2718, we carried out  HI interferometric observations of UGC 4703 using the GMRT.  The system was observed on June 15th, 2017. A baseband bandwidth of 16 MHz was used for the observations yielding a velocity resolution $\sim$7 km/s. The GMRT primary beam  at L band is 24$^{\prime}$ and the synthesised beam  of the images presented in the paper is 45.4$^{\prime\prime}$ $\times$ 38.0$^{\prime\prime}$. The data was analysed using the software {\tt AIPS}\footnote{http://www.aips.nrao.edu} and the procedure followed was similar to that explained in \cite{sengupta17}. Figure 2, top panel shows the contours from the integrated HI map overlaid on the SDSS $g$, $r$ and $i$-band combined image. While an HI bridge is visible validating an interaction between UGC 4703 and UGC 4703B, it is clear that there is no apparent sign of interaction between the MW type NGC 2718 and the dwarf satellites in both optical and radio HI images. In the interacting satellites, the HI emission is mainly concentrated around UGC 4703 with an extension of tenuous emission in the direction of UGC 4703B along the stellar bridge. In UGC 4703, the peak HI column density (1.6 $\times$ 10$^{20}$ cm$^{-2}$) overlaps with its bright central star-forming region. Interestingly, HI emission around  UGC 4703B does not overlap with the optical counterpart. The emission is not strong, the HI column density levels at UGC 4703B are similar to the bridge (between 2 - 6 $\times$ 10$^{19}$ cm$^{-2}$) with no prominent HI column density concentration on the galaxy. The integrated HI line flux density of the UGC 4703/ 4703B system estimated from our GMRT data is $\sim$ 1.4 Jy km/s compared to the uncorrected flux density estimated from the Arecibo spectrum from NED $\sim$ 2.0 Jy km/s. Two factors contributed to this flux discrepancy. Our data quality of that day was not good and heavy flagging resulted in losing crucial UV coverage. Additionally, the GMRT synthesised beam could have resolved out some diffuse emission from the extended, low column density bridge area of the UGC 4703 system. Thus the Arecibo flux density was used to estimate the total HI mass of the system, UGC 4703/ UGC 4703B combined, which is 1.4 $\times$ 10$^{9}$ M$_{\sun}$. Lack of better spatial resolution and the disturbed, interacting HI morphology forbids us from estimating more accurate individual HI masses of UGC 4703 and UGC 4703B. However, from the HI image and HI column density contours, we can conclude that gas content in UGC 4703B is significantly lower than in UGC 4703. It seems that such low gas mass content in UGC 4703B is intrinsic and which is not due to the detection limit of our observation. Figure 2 lower right panel shows the velocity field of the UGC 4703/ 4703B system. We detect hint of a modest velocity gradient in NGC 4703 which becomes irregular along the bridge and UGC 4703B  region though lack of better spatial resolution prevents us from making stronger claims about the gradient in UGC 4703.

\section{Discussion}\label{disc}

\subsection{Comparison with LMC-SMC-MW System}\label{lmc_comp}
Visible to the naked eye in southern sky, the LMC and SMC are the best-studied star-forming galaxies and thus the system is regarded as an important laboratory to study the evolution of dwarf galaxies. They are currently suffering different scales of tidal force from both the massive central host galaxy, MW, and their own lower mass companions. Understanding the origin of LMC-SMC system in vicinity of MW has been an active area of research and a widely discussed mechanism is  binary infall \citep{Besla07,Besla10,Diaz11,Kallivayalil09}. Particularly, \cite{Besla07} shows that LMC-SMC pair might be entering the MW halo for the first time and has not yet completed an orbit. This model was updated by \cite{Besla12} to explore the morphology of the stream produced from a head-on collision between the Clouds, specifically by the SMC moving in a highly eccentric orbit around the LMC, far from the MW potential. Here we report another system similar in geometry and morphology to LMC-SMC and compare their properties to gain a better understanding of the system with dwarf-dwarf interaction in proximity of a massive halo.

We list a direct comparison of physical parameters between the LMC-SMC-MW system and the UGC 4703 pair-NGC 2718 system in Table \ref{parcomb}. NGC 2718 and MW have a similar stellar mass of $\approx$5$\times$10$^{10}$ M$_{\sun}$. NGC 2718 is located in a fairly isolated environment. Its nearest bright (M$_{B}$ $< $-19 mag) galaxy is NGC 2731 at a sky-projected distance of 1.18 Mpc and relative line of sight velocity between the two is 1260 km/s. In contrast, MW is part of the Local group where the nearest bright neighbor is M33 at a distance of 0.86 Mpc.

Since we do not have resolved distances of each individual galaxies of the NGC  2718 group, the derived geometric properties are sky-projected. For the LMC-SMC-MW system, we use three-dimensional geometric properties which we obtained from \cite{Onghia16}. The geometrical configuration of both systems look similar.  The interacting dwarfs are located within a $\sim$100 kpc distance from the massive hosts, i.e NGC  2718 and MW, and the relative line of sight radial velocities between the hosts and dwarfs are $<$300km/s. The relative of line of sight velocity between UGC 4703 and UGC 4703B is 15 km/s and between LMC and SMC is 105 km/s.  Sky-projected separation between UGC 4703 and UGC 4703B  is 21 kpc and between LMC and SMC is 23 kpc.  The current star-formation rates of individual galaxies in UGC 4703 pair are also similar to the LMC-SMC pair. Stellar mass ratios of both interacting pairs,  UGC 4703-UGC 4703B and LMC-SMC, are  $\approx$5:1.

Although the tidal features of LMC-SMC pair is quite prominent in gas structure, recent observations have revealed presence of stellar substructure in the outskirts of the SMC \citep{Belokurov16,Belokurov17,Besla16}. The presence of a gaseous bridge between the LMC and SMC proves that  they must have had at least one close encounter in the recent past \cite{Kallivayalil13}. Additionally, \cite{Nidever13} identified  a $\approx$55 kpc stellar stream located to the east of SMC, a likely stellar counterpart of the HI Magellanic Bridge that was tidally stripped from the SMC \citep{Subramanian17}.  In comparison to the stellar streams observed in the LMC-SMC pair, the stellar streams around UGC 4703 pair is more prominent.  We also find a star-forming region (the putative TDG) in the stellar bridge which have a stellar mass log(M$_{*}$/M$_{\sun}$) = 7.25  and a star-formation rate 8.9$\times$10$^{-3}$ M$_{\sun}$/yr.  Overall, the stellar bridge is significantly bluer than both the dwarfs (UGC 4703, UGC 4703B) and is fairly aligned with the HI-bridge. This suggests, at least some, if not all, stellar population in the bridge may have formed very recently from the tidally stripped gas. Evidence of young stellar population has also been identified around the gaseous bridge connecting the LMC-SMC \citep{Belokurov16,Belokurov17}.

\begin{table}
\caption{Comparison with LMC-SMC-MW }
\begin{tabular}{cccccc}
\hline
Galaxy & M$_{B}$ & D & M$_{*}$ & SFR & HI$_{mass}$\\
 & mag & kpc & M$_{\sun}$ &M$_{\sun}$/yr & M$_{\sun}$\\
\hline
MW  & -21.17 & 0 & 5$\times$10$^{10}$ &  0.68 - 1.45 & 8$\times$10$^{9}$\\
LMC & -18.60 & 50 & 2.3$\times$10$^{9}$ & 0.2 & 4.4$\times$10$^{8}$\\
SMC & -17.20 & 60 & 5.3$\times$10$^{8}$ & 0.04 & 4$\times$10$^{8}$\\
\hline
NGC 2718 & -21.01 & 0 &  7$\times$10$^{10}$ &0.97 & 1.12$\times$10$^{10}$\\
UGC 4703  & -18.0 & 81 & 1.8 $\times$10$^{9}$ & 0.2 & 1.4 $\times$ 10$^{9*}$\\
UGC 4703B & -16.5 & 104 & 5.4$\times$10$^{8}$ & 0.03 &\\
\hline
\end{tabular}
\\
D is distance from the host galaxy to the dwarfs. In our case, it is sky-projected while the values for LMC and SMC are 3-dimensional. The LMC and SMC parameters are obtained from \cite{Onghia16}. $^{*}$Combined HI mass of UGC 4703 and UGC 4703B.
\label{parcomb}
\end{table}

As far as our detection limit permits no sign of an extended HI tail is observed around the UGC 4703-pair unlike the LMC-SMC system where a long trailing gaseous stream (well known as Magellanic Stream MS) is prominent. Origin of the MS is a debate and a frequent, but not conclusive, explanation is that it is a product of  tidal interactions between either LMC-SMC-MW or LMC-SMC only \citep{Fujimoto76,Guglielmo14}. There are other scenarios like ram-pressure stripping, that could also explain the origin of MS \citep{Meurer85,Moore94} if the interacting pair had already crossed the MW outer disk at some point in their past orbit. Even in a purely tidal interaction model, it is not clear how much MW potential affect the dynamical history of the pair or, in other words, whether the pair is bound to MW or not. Lack of three dimensional distance of NGC 2718 and UGC 4703 pair, prevents us from estimating if the UGC 4703 and UGC 4703B are bound to NGC 2718 or not.

We find that UGC 4703B is relatively gas poor and the HI map shows that the gas distribution is displaced from its optical counterpart while both LMC and SMC relatively are gas rich. This maybe a hint that the interaction between UGC 4703 and UGC 4703B is relatively advanced compare to LMC-SMC. Being the minor companion, UGC 4703B has experienced tidal stretching from UGC 4703 which probably has displaced most of the  gas mass and deformed the stellar disk creating a stellar bridge.  One main difference in geometry of LMC-SMC and UGC 4703-UGC 4703B is relative line of sight velocity between the pair, where the later has significantly lower value, i.e 15 km/s. In contrast to the previous hypothesis, this may also suggest that the encounter in UGC 4703-UGC 4703B is slower than encounter in LMC-SMC and that leads to a stronger tidal distortion of participating galaxies in UGC 4703-UGC 4703B compared to relatively fast interacting LMC-SMC. However, a comparative study with a result of numerical simulation with full orbital histories of interactions is required to make a firm conclusion.

In summary, we present UGC 4703 pair- NGC 2718 system as a LMC-SMC-MW analog. Both the systems have a similar physical (geometry, star-formation rate, total gas mass and stellar mass) and morphological properties. Our GMRT observations detected HI in  NGC 2718  and UGC 4703 pair as well as in the bridge between the dwarf pair. However no extended HI, similar to the MS, is detected between NGC 2718  and UGC 4703 pair. We also detected star-forming regions along the UGC 4703/ 4703B bridge with stellar mass exceeding 10$^{7}$ M$_{\sun}$. A comparison of optical and HI morphology of interacting dwarfs pairs (UGC 4703-4703B and LMC-SMC) suggests interaction between UGC 4703 and UGC 4703B is either {\it slow or at relatively advanced stage} compare to LMC-SMC interaction.

\acknowledgments
Paudel S. acknowledges the support by Samsung Science \& Technology Foundation under Project Number SSTF-BA1501-0. We thank the staff of the GMRT that made these observations possible. GMRT is run by the National Centre for Radio Astrophysics of the Tata Institute of Fundamental Research. This work is based archival data from the SDSS-III, Spitzer telescope and GALEX.


\end{document}